\documentclass[12pt]{article}
\usepackage{axodraw}
\newcommand{\beq}{\begin{equation}}
\newcommand{\eeq}{\end{equation}}
\newcommand{\beqa}{\begin{eqnarray}}
\newcommand{\eeqa}{\end{eqnarray}}
\newcommand{\beqar}{\begin{eqnarray*}}
\newcommand{\eeqar}{\end{eqnarray*}}

\def\non          {\nonumber}
\def\ha           {\mbox{$\frac{1}{2}$}}

\def\Tr           {\mbox{\rm Tr}\,}
\def\STr          {\mbox{\rm STr}\,}

\def\fsH	{H\!\!\!\!/\,}

\newcommand{\eps}{\epsilon}
\newcommand{\ga}{\gamma}
\newcommand{\Ga}{\Gamma}

\newcommand{\inn}{\!\cdot\!}

\renewcommand{\l}{\lambda}

\newcommand{\z}{\zeta}

\newcommand{\eg}{{\it e.g.,}\ }
\newcommand{\ie}{{\it i.e.,}\ }
\newcommand{\labell}[1]{\label{#1}} %{\label{#1}} %
\newcommand{\reef}[1]{(\ref{#1})}
\newcommand\prt{\partial}

\newcommand\bD{\bar{D}}

\begin{document}
\baselineskip 18pt%
\begin{titlepage}
\vspace*{1mm}%
\hfill%
\vspace*{15mm}%

\centerline{{\Large {\bf Higher derivative corrections to    Wess-Zumino  }}}\vspace*{3mm} \centerline{{\Large {\bf     action of Brane-Antibrane systems }}}
\vspace*{5mm}
\begin{center}
{ Mohammad R. Garousi}%

\vspace*{0.8cm}{ {Department of Physics, Ferdowsi university,
P.O. Box 1436, Mashhad, Iran}}\\
%{\it {and}}\\
{ { Institute for Studies in Theoretical Physics and Mathematics (IPM) \\
P.O.Box 19395-5531, Tehran, Iran}}\\
%{E-mails: {\tt garousi@mail.ipm.ir, eh$_{-}$ha420@ stu-mail.um.ac.ir\\
%\quad\quad\quad\quad\quad\quad\quad\quad\quad\quad,  ehsanhatefi@gmail.com}}\\
\vspace*{1.5cm}
\end{center}

\begin{center}{\bf Abstract}\end{center}
\begin{quote}
By explicit calculation, we show that the  expansion of the disk level S-matrix element of one  RR field, two open string tachyons  and one gauge field that has been recently found  corresponds to the derivative expansion of the Wess-Zumino action of D-brane-anti-D-brane systems.

\end{quote}
\end{titlepage}%--------------------------------------------------------------------
\section{Introduction}
Study of unstable objects in string theory  might shed new light
in understanding properties of string theory in time-dependent
backgrounds \cite{Gutperle:2002ai,Sen:2002in,Sen:2002an,Sen:2002vv,Lambert:2003zr,Sen:2004nf,Ohta:2003uw}. Generally
speaking,  source of instability in these processes  is appearance
of some tachyonic  modes  in the spectrum of these 
objects. It  then makes sense to study them in a field
theory which includes those modes. In this regard, it has been
shown by A. Sen that an effective action of the Dirac- Born-Infeld type
proposed in \cite{Sen:1999md,Garousi:2000tr,Bergshoeff:2000dq,Kluson:2000iy} can capture many properties
of the decay of non-BPS D$_p$-branes in string theory
\cite{Sen:2002in,Sen:2002an}. This action has been found in \cite{Garousi:2000tr} by studying the  S-matrix element of one graviton and two tachyons. 
%Having an effective action, one may then
%study   evolution of these unstable objects in time-dependent
%backgrounds. See \cite{Tachyonindustry} for possible cosmological
%application of the tachyonic DBI action.

Recently, unstable objects have been used to study spontaneous chiral symmetry breaking  in  holographic model of QCD \cite{Casero:2007ae,Bergman:2007pm,Dhar:2007bz}. In these studies, flavor branes introduced by placing a set of parallel branes and antibranes on a background dual to a confining color theory \cite{Sakai:2004cn}. 
%When the number of flavour branes are much much smaller than the number of color branes, the branes-antibranes  may be %considered as a prob.  
Detailed study of brane-anti-brane system
 reveals   when branes separation is smaller than the
string length scale,  the spectrum  has two tachyonic
modes \cite{Sen:1998ii}.  The
 effective action  should then include
these  modes as they are the most important ones
which rule  the dynamics of   the system.

The effective action of a $D_p\bD_p$-brane in Type IIA(B) theory should be  given by some extension of  the DBI action and the WZ terms which include the tachyon fields. The DBI part may be given by the projection of the effective action of two non-BPS $D_p$-branes in Type IIB(A) theory with $(-1)^{F_L}$ projection \cite{Garousi:2004rd}. We are interested in this paper in the appearance of tachyon, gauge field and the RR field in these actions. These fields appear in  the DBI part as the following \cite{Garousi:2007fn}:  
\beqa
S_{DBI}&=&-T_p\int
d^{p+1}\sigma \Tr\left(V({\cal T})
\sqrt{-\det(\eta_{ab}
+2\pi\alpha'F_{ab}+2\pi\alpha'D_a{\cal T}D_b{\cal T})} \right)\,\,,\labell{nonab} \eeqa where $T_{p}$ is the
p-brane tension. The trace in the above action 
should be completely symmetric between all  matrices
of the form $F_{ab},D_a{\cal T}$, and individual
${\cal T}$ of the tachyon potential.   These matrices  are
\beqa
F_{ab}=\pmatrix{F^{(1)}_{ab}&0\cr 
0&F^{(2)}_{ab}},\,\,
D_{a}{\cal T}=\pmatrix{0&D_aT\cr 
(D_aT)^*&0},\,\, {\cal T}=\pmatrix{0&T\cr 
T^*&0}\,\labell{M12} \eeqa 
where $F^{(i)}_{ab}=\prt_{a}A^{(i)}_{b}-\prt_{b}A^{(i)}_{a}$ and $D_{a}T=\prt_{a}T-i(A^{(1)}_a-A^{(2)}_a)T$.  
The tachyon potential which is consistent with S-matrix element calculations
has the following expansion: \beqa
V({\cal T})&=&1+\pi\alpha'm^2{\cal T}^2+
\frac{1}{2}(\pi\alpha'm^2{\cal T}^2)^2+\cdots
\non\eeqa  where $m^2$ is the mass squared of tachyon, \ie
$m^2=-1/(2\alpha')$. The above expansion is consistent with the
potential $V({\cal T})=e^{\pi\alpha'm^2{\cal T}^2}$ which is the tachyon
potential of BSFT ~\cite{Kutasov:2000aq}.
 This action  has
 the following expansion: 
\beqa {\cal
L}_{DBI}&\!\!\!=\!\!\!&-2T_p-T_p(2\pi\alpha')\left(m^2|T|^2+DT\cdot(DT)^{*}-\frac{\pi\alpha'}{2}
\left(F^{(1)}\cdot{F^{(1)}}+
F^{(2)}\cdot{F^{(2)}}\right)\right)+\cdots\labell{exp1}
\eeqa
where dots refers  to the terms which have more than two fields. 

The  WZ 
term describing the  coupling of RR field to tachyon and gauge field of brane-anti-brane is given by \cite{Kennedy:1999nn,Kraus:2000nj,Takayanagi:2000rz}\beqa
S_{WZ}&=&\mu_p \int_{\Sigma_{(p+1)}} C \wedge \STr e^{i2\pi\alpha'\cal F}\labell{WZ}\eeqa 
where the curvature of the superconnection is defined as:
\beqa {\cal F}&=&d{\cal A}-i{\cal A}\wedge\cal A\eeqa
the superconnection is \begin{displaymath}
i{\cal A} = \left(
\begin{array}{cc}
  iA^{(1)} & \beta T^* \\ \beta T &   iA^{(2)} 
\end{array}
\right) \ ,
\non\end{displaymath}
where $\beta$ is a normalization constant. If one uses the tachyon DBI action \reef{nonab} for describing the dynamics  of the tachyon  field then the normalization of tachyon in the WZ action  \reef{WZ} has to be  \cite{Garousi:2007fk}
\beqa
\beta&=&\frac{1}{\pi} \sqrt{\frac{2\ln(2)}{\alpha'}}\eeqa
The ``supertrace'' in \reef{WZ} is defined by
\begin{displaymath}
\STr \left( \begin{array}{cc} A&B\\C&D \end{array} \right)
= \Tr A - \Tr D \ .
\non\end{displaymath} Using the multiplication rule of the supermatrices \cite{Kraus:2000nj}
\beqa
 \left
( \begin{array}{cc}A&B\\C&D\end{array} 
\right)\cdot \left( \begin{array}{cc}A'&B'\\C'&D'\end{array} 
\right)\,=\,\left( \begin{array}{cc}AA'+(-)^{c'}BC'&AB'+(-)^{d'}BD'\\DC'+(-)^{a'}CA'&DD'+(-)^{b'}CB'\end{array} 
\right)\eeqa where $x'$ is 0 if $X$ is an even form or 1 if $X$ is an odd form, one finds that the curvature  is
\begin{displaymath}
i{\cal F} = \left(
\begin{array}{cc}
iF^{(1)} -\beta^2 |T|^2 & \beta (DT)^* \\
\beta DT & iF^{(2)} -\beta^2|T|^2 
\end{array}
\right) \ ,
\non\end{displaymath}
where $F^{(i)}=\frac{1}{2}F^{(i)}_{ab}dx^{a}\wedge dx^{b}$ and $DT=[\partial_a T-i(A^{(1)}_{a}-A^{(2)}_{a})T]dx^{a}$. Using the expansion for the exponential term in the WZ action \reef{WZ}, one finds many different terms. The terms which involve at most three open string fields are the following:
\beqa
\mu_p(2\pi\alpha')C\wedge \STr i{\cal F}&\!\!\!\!=\!\!\!&\mu_p(2\pi\alpha')C_{p-1}\wedge(F^{(1)}-F^{(2)})\labell{exp2}\\
\frac{\mu_p}{2!}(2\pi\alpha')^2C\wedge \STr i{\cal F}\wedge i{\cal F}&\!\!\!\!=\!\!\!\!&\frac{\mu_p}{2!}(2\pi\alpha')^2C_{p-3}\wedge \left\{F^{(1)}\wedge F^{(1)}-
F^{(2)}\wedge F^{(2)}\right\}\nonumber\\
&& +C_{p-1}\wedge\left\{-2\beta^2|T|^2(F^{(1)}-F^{(2)})+2i\beta^2 DT\wedge(DT)^*\right\}\nonumber\\
\frac{\mu_p}{3!}(2\pi\alpha')^3C\wedge \STr i{\cal F}\wedge i{\cal F}\wedge i{\cal F}&\!\!\!\!=\!\!\!\!&
\frac{\mu_p}{3!}(2\pi\alpha')^3C_{p-3}\left\{3i\beta^2(F^{(1)}+F^{(2)})\wedge DT\wedge(DT)^*\right\}\nonumber
\eeqa
  The coupling of one RR field $C_{p-1}$, two tachyons and one gauge field in the above terms  can be combined into the following form:
\beqa
&-\beta^2\mu_p(2\pi\alpha')^2 \int_{\Sigma_{(p+1)}} C_{(p-1)}\wedge \left\{d(A^{(1)}-A^{(2)})TT^*-(A^{(1)}-A^{(2)})d(TT^*)\right\}
&\nonumber\\
&=-\beta^2\mu_p(2\pi\alpha')^2\int_{\Sigma_{p+1}}H_{(p)}\wedge (A^{(1)}-A^{(2)})TT^*&\labell{contact}\eeqa
This combination actually appears naturally in the S-matrix element in the string theory side \cite{Garousi:2007fk}. It has been shown in \cite{Garousi:2007fk} that the effective actions \reef{nonab} and \reef{WZ} are consistent with an expansion of the S-matrix element of one RR, two tachyons and one gauge field. In the present paper, we would like to show that the expansion found in \cite{Garousi:2007fk} is in fact the momentum expansion, \ie the expansion which is consistent with the derivative expansion of the field theory  of brane-anti-brane system.  We will show this by explicitly calculating the higher derivative terms of the WZ field theory, \ie \reef{hderv12}, \reef{hderv2}, and \reef{highpn'}. 

An outline of the rest of paper is as follows. In the next section,  we study the momentum expansion  of the S-matrix element of one RR and two tachyons, and the S-matrix element of one RR and two gauge fields. We shall find in this section the higher derivative extension of the coupling in the second line of \reef{exp2} and the coupling in the last term in the third line of \reef{exp2}.  In section 3, we study the momentum  expansion of  the S-matrix element of one RR, two tachyons and one gauge field. We shall find the higher derivatives extension of the coupling in the last line of \reef{exp2} and the coupling in the first term in the third line of \reef{exp2}. In this section we will also find a class of higher derivative terms, \ie  \reef{highpn'} which has at least four derivatives, hence, they are not higher derivative extension of the two derivative couplings of the WZ terms. We discuss briefly our results in section 4, and give a  general rule for finding the  momentum  expansion of any S-matrix element involving the tachyon fields.

%%%%%%%%%%%%%%%%%%%%%%%%%%%%%%%%%%%%%%%%%%%%%%%%%%%%%%%%%%%%%%%%%%%%%%%%%%%%%

 %%%%%%%%%%%%%%%%%%%%%%%%%%%%%%%%%%%%%%%%%%%%%%%%%%%%%%%%%%%%%%%%%%%%%%%%%%%%%%%%%%%%%%%
\section{Three-point function}

The three-point amplitude between one RR field and two  tachyons in string theory side is given as \cite{Kennedy:1999nn,Garousi:2007fk}  
\beqa
{\cal A}^{T,T,RR}
&=&\left(\frac{i\mu_p}{4}\right)2\pi \frac{\Ga[-2u]}{\Ga[\ha-u]^2}
 \Tr (P_{-}\fsH_{(n)}M_p\ga^{a})k_{a} \labell{amp33}\ .
\eeqa
where $u = -(k+k')^2$ and $k,\, k'$ are the momenta of the tachyons. In the string theory side we have set $\alpha'=2$. The trace is zero for $p\neq n$, and for $n=p$ it is
\begin{displaymath}
\Tr \left( \fsH_{(n)}M_p\ga^{a}\right)
= \pm\frac{ 32}{p!}H_{a_{0}\ldots a_{p-1}}
 \eps^{a_{0}\ldots a_{p-1}a} \ .
\non\end{displaymath}
We are going to compare string theory S-matrix elements with field theory S-matrix element including their coefficients, however, we are not interested in fixing the overall sign of the amplitudes. Hence, in above and in the rest of equations in this paper, we have payed   no attention to the sign of  equations. The trace in \reef{amp33} containing the factor of $\ga^{11}$ ensures the following
results also hold for $p>3$ with $H_{(n)} \equiv \ast H_{(10-n)}$ for
$n\geq 5$. The tachyon vertex operator in string theory corresponds to the real  components of the complex tachyon of field theory, \ie
\beqa
T&=&\frac{1}{\sqrt{2}}(T_1+iT_2)\eeqa
Now if one  replaces $k_a$ in \reef{amp33} with $-k'_a-p_a$ using  the conservation of momentum,  one will find  that the $p_a$ term vanishes using the totally antisymmetric property of $\eps^{a_{0}\ldots a_{p-1}a}$. Hence the amplitude \reef{amp33} is antisymmetric under interchanging $1\leftrightarrow 2$. This indicates that  only  the three-point amplitude between one RR, one  $T_1$ and one $T_2$ is non-zero.

The momentum  expansion of \reef{amp33} is at $u\rightarrow 0$. Using the Maple,  one can expand the
prefactor of \reef{amp33} around this point, \ie  
\beqa
2\pi\frac{\Ga[-2u]}{\Ga[\ha-u]^2}
 &=&  \frac{-1}{
 u} + \sum_{n=0}^{\infty}a_nu^u
\ .\labell{taylor}
\eeqa
where some of the coefficients $a_n$ are 
\beqa
a_0&=&4\ln(2)\nonumber\\
a_1&=&\frac{\pi^2}{6}-8\ln(2)^2\nonumber\\
a_2&=&-\frac{2}{3}\left(\pi^2\ln(2)-3\z(3)-16\ln(2)^3\right)\labell{as}\\
a_3&=&\frac{1}{120}\left(160\pi^2\ln(2)^2+3\pi^4-960\z(3)\ln(2)-1280\ln(2)^4\right)\nonumber\\
a_5&=&-\frac{1}{90}\left(160\pi^2\ln(2)^3+9\pi^4\ln(2)-1440\z(3)\ln(2)^2+30\z(3)\pi^2-768\ln(2)^5-540\z(5)\right)\nonumber
\eeqa
It is shown in \cite{Kennedy:1999nn} that the massless pole reproduced by the kinetic term of tachyon and the WZ coupling in the first line of \reef{exp2}.  There is no higher power of momenta in the massless pole, hence, the kinetic term of tachyon and the WZ coupling  have no higher derivative extension. 
Since the expansion \reef{taylor} is in terms of the powers of $p_a^2$, the other  terms in \reef{taylor}  correspond to the higher derivative corrections of  the WZ action.  It is easy to check that the following  higher derivative terms  reproduce the other terms  in \reef{taylor}:
\beqa
2i\alpha'\mu_p\sum_{n=0}^{\infty}a_n\left(\frac{\alpha'}{2}\right)^n C_{p-1}\wedge (D ^aD_a)^n(DT\wedge DT^*)\labell{hderv}
\eeqa
The above couplings have on-shell ambiguity, since one can replace $T$ with $\prt^a\prt_a T$ for the on-shell external tachyon. However, we will show in the next section that the above couplings, with exactly the same coefficients $a_n$, appear in the contact terms as well as in the tachyonic pole of the scattering amplitude of one RR, two tachyons and one gauge field. This indicates that there is no on-shell ambiguity in the above couplings. We shall discus it more in the Discussion section. So, the above couplings are the higher derivative extension of the coupling in the second term in the third line of \reef{exp2}.  
 
 The string theory S-matrix element of one RR and two gauge fields is given by \cite{Hashimoto:1996kf,Garousi:1998fg}
\beqa
{\cal A}&\sim&2\frac{\Gamma[-2u]}{\Gamma[1-u]^2}K\eeqa
where $K$ is the kinematic factor. Obviously the momentum  expansion of this amplitude is around $u\rightarrow 0$. Expansion of the prefactor at this point is
\beqa
2\frac{\Ga[-2u]}{\Ga[1-u]^2}
 &=&    -\sum_{n=-1}^{\infty}b_nu^n 
\ .\labell{taylor1}
\eeqa
where some of the coefficients $b_n$ are
\beqa
b_{-1}&=&1\nonumber\\
b_0&=&0\nonumber\\
b_1&=&\frac{\pi^2}{6}\nonumber\\
b_2&=&2\z(3)\nonumber\\
b_3&=&\frac{19}{360}\pi^4\labell{bs}\\
b_4&=&\frac{1}{3}\left(\z(3)\pi^2+18\z(5)\right)\nonumber\\
b_5&=&\frac{1}{3024}\left(55\pi^6+6048\z(3)^2\right)\nonumber\eeqa
In this case actually there is no massless pole at $u=0$ as the  kinematic factor  provides a compensating factor of $u$. The amplitude  has the following expansion:
\beqa
{\cal A}&=&i\frac{(4\pi)^2\mu_p}{4(p-3)!}f^{a_0a_1}{f'}^{a_2a_3}\varepsilon^{a_4\cdots a_p}\epsilon_{a_0\cdots a_p}\left(\sum_{n=-1}^{\infty}b_nu^{n+1}\right)\delta_{p,n+2}
\eeqa
where $f_{ab}=i(k_a\xi_b-k_b\xi_a)$, ${f'}_{ab}=i(k'_a\xi'_b-k'_b\xi'_a)$ and $\varepsilon$ is the polarization of the RR potential. The above terms are  reproduced   by the following higher derivative couplings of field theory
\beqa
\frac{\mu_p}{2!}(2\pi\alpha')^2C_{p-3}\wedge \left(\sum_{n=-1}^{\infty}b_n(\alpha')^{n+1}\partial^{a_1}\cdots\partial^{a_{n+1}} F^{(1)}\wedge \partial_{a_1}\cdots\partial_{a_{n+1}} F^{(1)}-\left(F^{(1)}\rightarrow F^{(2)}\right)\right)\labell{highaa}\eeqa
We will see in the next section that these couplings, with exactly the same coefficients $b_n$,  appear  in the massless pole of the scattering amplitude of one RR, two tachyons and one gauge field. The above couplings are the higher derivative extension of the coupling in the second line of \reef{exp2}.
 %%%%%%%%%%%%%%%%%%%%%%%%%%%%%%%%%%%%%%%%%%%%%%%%%%%%%%%%%%%%%%%%%%%%%%%%%%%%%%%%%%%%%%%

\section{Four-point function}
The S-matrix element of  one RR field,  two tachyons   and 
one gauge field is given as \cite{Garousi:2007fk}
\beqa
{\cal A}^{ATTC}&\!\!\!\!\!=\!\!\!\!\!&\frac{i\mu_p}{2\sqrt{2\pi}}\left[\Tr\bigg((P_{-}\fsH_{(n)}M_p)
(k_3.\ga)(k_2.\ga)(\xi.\ga)\bigg)I\delta_{p,n+2}
+\Tr\bigg((P_{-}\fsH_{(n)}M_p)
\ga^{a}\bigg)J\delta_{p,n}\right. \nonumber\\&&\left.\times
\bigg\{k_{2a}(t+1/4)(2\xi.k_{3})
+k_{3a}(s+1/4)(2\xi.k_{2})-\xi_a(s+1/4)(t+1/4)\bigg\}\right]\labell{gen}\eeqa
where   $I,\,J$ are :
 \beqa I&=&2^{1/2}(2)^{-2(t+s+u)-1}\pi{\frac{\Gamma(-u)
 \Gamma(-s+1/4)\Gamma(-t+1/4)\Gamma(-t-s-u)}{\Gamma(-u-t+1/4)
\Gamma(-t-s+1/2)\Gamma(-s-u+1/4)}}\nonumber\eeqa
\beqa J&=&2^{1/2}(2)^{-2(t+s+u+1)}\pi{\frac{\Gamma(-u+1/2)
\Gamma(-s-1/4)\Gamma(-t-1/4)\Gamma(-t-s-u-1/2)}
{\Gamma(-u-t+1/4)\Gamma(-t-s+1/2)\Gamma(-s-u+1/4)}}\nonumber\eeqa 
where   the Mandelstam variables are\beqar
s&=&-(k_1+k_3)^2,\qquad t=-(k_1+k_2)^2,\qquad u=-(k_2+k_3)^2
\qquad\eeqar $k_1$ is momentum of the gauge field and $k_2,\, k_3$ are the momenta of the tachyons. Note that  $I$, $J$ are symmetric under $s\leftrightarrow t$. The  traces in \reef{gen} are: \beqa
\Tr\bigg(\fsH_{(n)}M_p
(k_3.\ga)(k_2.\ga)(\xi.\ga)\bigg)\delta_{p,n+2}&=&\pm\frac{32}{n!}\eps^{a_{0}\cdots a_{p}}H_{a_{0}\cdots a_{p-3}}
k_{3a_{p-2}}k_{2a_{p-1}}\xi_{a_p}\delta_{p,n+2}\nonumber\\
\Tr\bigg(\fsH_{(n)}M_p
\ga^a\bigg)\delta_{p,n}&=&\pm\frac{32}{n!}\eps^{a_{0}\cdots a_{p-1}a}H_{a_{0}\cdots a_{p-1}}
\delta_{p,n}
\eeqa
Examining  the poles of the Gamma functions, one realizes that for the case that $p=n+2$, the amplitude has massless pole and infinite tower of massive poles. Whereas for $p=n$ case, there are tachyon, massless, and infinite tower of massive poles. The tachyon pole in particular indicates that the kinetic term of the tachyon  has no higher derivative extension. It has been shown in \cite{Garousi:2007fk} that the leading order term of the amplitude \reef{gen} expanded around the following point :
\beqa t\rightarrow{-1/4},\qquad s\rightarrow{-1/4},\qquad
u\rightarrow{0} \labell{point}\eeqa 
is consistent with the effective actions \reef{nonab} and \reef{WZ}. We would like to find the field theory couplings which reproduce all terms of the expansion.  Let us study each case separately. 

\subsection{$p=n+2$ case}

For $p=n+2$, the amplitude is antisymmetric under interchanging $ 2 \leftrightarrow 3$, hence the four-point function between one RR, one gauge field and two $T_1$ or two $T_2$ is zero. The electric part of the amplitude for one RR, one gauge field, one $T_1$ and one $T_2$ is given by 
\beqa
{\cal A}^{AT_1T_2C}&=&\pm\frac{8i\mu_p}{\sqrt{2\pi}(p-2)!}\left[ \eps^{a_{0}\cdots a_{p}}H_{a_{0}\cdots a_{p-3}}
k_{3a_{p-2}}k_{2a_{p-1}}\xi_{a_p}\right]I\labell{pn2}\eeqa 
Note that the amplitude satisfies the Ward identity, \ie the amplitude vanishes under replacement $\xi^a\rightarrow k_1^a$. 

Expansion of $I$ around \reef{point} is
\beqa
I&=&
\pi\sqrt{2\pi}\left(-\frac{1}{u}\sum_{n=-1}^{\infty}b_n(s+t+1/2)^{n+1}
+\right.\labell{expI}\\
&&\left.+\sum_{p,n,m=0}^{\infty}c_{p,n,m}u^p\left((s+1/4)(t+1/4)\right)^n(s+t+1/2)^m\right)\nonumber\eeqa
where the coefficients $b_n$ are exactly those that appear in \reef{bs} and  $c_{p,0,0}=a_p$ are those that appear in \reef{as}. The  constants $c_{p,n,m}$ for some other cases are the following:
\beqa
&&c_{0,0,2}=\frac{2}{3}\pi^2\ln(2),\,c_{0,1,0}=-14\z(3),
c_{0,0,3}=8\z(3)\ln(2),\,\\
&&c_{1,1,0}=56\z(3)\ln(2)-1/2,\,c_{1,0,2}=\frac{1}{36}(\pi^4-48\pi^2\ln(2)^2),\,c_{0,1,1}=-1/2\nonumber
\eeqa
Inserting the first term of \reef{expI} into \reef{pn2}, one finds a massless pole which must be reproduced by field theory couplings.

The couplings in  \reef{exp1} and \reef{highaa},  produces 
 the following massless pole for $p=n+2$:
\beqa
{\cal A}&=&V_a(C_{p-3},A^{(1)},A^{(1)})G_{ab}(A^{(1)})V_b(A^{(1)},T_1,T_2)\labell{amp3}\eeqa
where
\beqa
G_{ab}(A^{(1)}) &=&\frac{i\delta_{ab}}{(2\pi\alpha')^2 T_p
\left(u\right)}\nonumber\\
V_b(A^{(1)},T_1,T_2)&=&T_p(2\pi\alpha')(k_2-k_3)_b\\
V_a(C_{p-3},A^{(1)},A^{(1)})&=&\mu_p(2\pi\alpha')^2\frac{1}{(p-2)!}\epsilon_{a_0\cdots a_{p-1}a}H^{a_0\cdots a_{p-3}}k_1^{a_{p-2}}\xi^{a_{p-1}}\sum_{n=-1}^{\infty}b_n(\alpha'k_1\cdot k)^{n+1}\nonumber\eeqa
where $k$ is the momentum of the off-shell gauge field. Note that the vertex $V_b(A^{(1)},T_1,T_2)$ has no higher derivative correction as it arises from the kinetic term of the tachyon.  The amplitude \reef{amp3} becomes
\beqa
{\cal A}=\mu_p(2\pi\alpha')\frac{2i}{(p-2)!u}\epsilon_{a_0\cdots a_{p-1}a}H^{a_0\cdots a_{p-3}}k_2^{a_{p-2}}k_3^{a_{p-1}}\xi^a\sum_{n=-1}^{\infty}b_n\left(\frac{\alpha'}{2}\right)^{n+1}(s+t+1/2)^{n+1}\labell{AA}\eeqa
this is exactly the massless pole of string theory amplitude. Note that there is no left over residual contact term in comparing above amplitude with the massless pole of the string theory amplitude.

The contact terms of string theory amplitude \reef{expI} on the other hand are reproduced by the following couplings:
\beqa
&&2i\alpha'(\pi\alpha')\mu_p\sum_{p,n,m=0}^{\infty}c_{p,n,m}\left(\frac{\alpha'}{2}\right)^{p}\left(\alpha'\right)^{2n+m} C_{p-3}\wedge \prt^{a_1}\cdots\prt^{a_{2n}}\prt^{b_1}\cdots\prt^{b_{m}}(F^{(1)}+F^{(2)})\nonumber\\
&&\wedge(D^aD_a)^p  D_{b_1}\cdots D_{b_{m}}(D_{a_1}\cdots D_{a_n}DT\wedge D_{a_{n+1}}\cdots D_{a_{2n}}DT^*)\labell{hderv12}
\eeqa
For $n=m=0$ case, the above couplings are  the natural extension of the couplings \reef{hderv} to $C_{p-3}$. Since there is no on-shell ambiguity for the couplings in \reef{hderv}, one expects there should be no on-shell ambiguity  for the above couplings either.  The above couplings are the higher derivative extension of the coupling in the fourth line of \reef{exp2}.

\subsection{$p=n$ case}

Now we consider  $n=p$ case. The string theory amplitude in this case is symmetric under interchanging $2 \leftrightarrow 3$. On the other hand, there is no Feynman amplitude in field theory corresponding to four-point function of one RR, one gauge field, one $T_1$ and one $T_2$. Hence, for $p=n$ the string theory amplitude \reef{gen} is the S-matrix element of one RR, one gauge field and two $T_1$ or two $T_2$. Its electric part is,
\beqa
{\cal A}^{AT_1T_1C}&\!\!\!\!=\!\!\!\!&\pm\frac{8i\mu_p}{\sqrt{2\pi}p!}\left[\bigg( \eps^{a_{0}\cdots a_{p-1}a}H_{a_{0}\cdots a_{p-1}}\bigg)J\right. \nonumber\\&&\left.\times
\bigg\{k_{2a}(t+1/4)(2\xi.k_{3})
+k_{3a}(s+1/4)(2\xi.k_{2})-\xi_a(s+1/4)(t+1/4)\bigg\}\right]\labell{pn}\eeqa
%where we have also multiply the amplitude by the factor of 2 since amplitude is symmetric under $2\leftrightarrow 3$.
 Note that the amplitude satisfies the Ward identity, \ie the amplitude vanishes under replacement $\xi^a\rightarrow k_1^a$. 

The expansion of $(s+1/4)(t+1/4)J$  around \reef{point} is
\beqa
&&(s+1/4)(t+1/4)J=\frac{\sqrt{2\pi}}{2}\left(\frac{-1}{(t+s+u+1/2)}+ \sum_{n=0}^{\infty}a_n(s+t+u+1/2)^n\right.\nonumber\\
&&\left.+\frac{\sum_{n,m=0}^{\infty}d_{n,m}(s+t+1/2)^n((t+1/4)(s+1/4))^{m+1}}{(t+s+u+1/2)}\right.\labell{high}\\
&&\left.+\sum_{p,n,m=0}^{\infty}e_{p,n,m}(s+t+u+1/2)^p(s+t+1/2)^n((t+1/4)(s+1/4))^{m+1}\right)\nonumber\eeqa
where the coefficients $a_n$ in the first line are exactly those appear in \reef{as}. Some of the coefficients $d_{n,m}$ and $c_{p,n,m}$ are\beqa
d_{0,0}=-\pi^2/3,\,&&d_{1,0}=8\z(3)\\
d_{2,0}=-7\pi^4/45,\, d_{0,1}=\pi^4/45,\,&&\,d_{3,0}=32\z(5),\, d_{1,1}=-32\z(5)+8\z(3)\pi^2/3\nonumber\\
e_{0,0,0}=\frac{2}{3}\left(2\pi^2\ln(2)-21\z(3)\right),&& e_{1,0,0}=\frac{1}{9}\left(4\pi^4-504\z(3)\ln(2)+24\pi^2\ln(2)^2\right)\nonumber
\eeqa
Note that the contact terms in the last line of \reef{high} do not have the structure of   the contact terms in the first line of \reef{high}. They correspond to different couplings in field theory. The  field theory,  has the following massless poles for $p=n$:
\beqa
{\cal A}&=&V_a(C_{p-1},A)G_{ab}(A)V_b(A,T_1,T_1,A^{(1)})\labell{amp4}\eeqa
where  $A$ should  be $A^{(1)}$ and $A^{(2)}$. The propagator and vertexes  $ V_a(C_{p-1},A)$  are
\beqa
G_{ab}(A) &=&\frac{i\delta_{ab}}{(2\pi\alpha')^2 T_p
\left(u+t+s+1/2\right)}\nonumber\\
V_a(C_{p-1},A^{(1)})&=&i\mu_p(2\pi\alpha')\frac{1}{p!}\epsilon_{a_0\cdots a_{p-1}a}H^{a_0\cdots a_{p-1}}\labell{amp4'}\\
V_a(C_{p-1},A^{(2)})&=&-i\mu_p(2\pi\alpha')\frac{1}{p!}\epsilon_{a_0\cdots a_{p-1}a}H^{a_0\cdots a_{p-1}}\nonumber
\eeqa
If one uses the kinetic term of the tachyon   to find the vertex $V_b(A^{(1)},T_1,T_1,A^{(1)})$
then the amplitude  \reef{amp4}  reproduces the massless pole in the first term of \reef{high}.  To find the higher derivative coupling corresponding to the second term in \reef{high}, we consider  the following higher derivative terms: 
\beqa
-2\alpha'\mu_p\sum_{n=0}^{\infty}a_n\left(\frac{\alpha'}{2}\right)^n C_{p-1}\wedge (D^aD_a)^n[(F^{(1)}-F^{(2)})|T|^2]\labell{hderv2}
\eeqa
Combining  the above with  the  coupling of one RR, two tachyons and one gauge field  of \reef{hderv}, one finds the following coupling:
\beqa
-2\alpha'\mu_p\sum_{n=0}^{\infty}a_n \left(\frac{\alpha'}{2}\right)^n H_p\wedge (\partial^a\partial_a)^n[(A^{(1)}-A^{(2)})TT^*]\labell{cont1}\eeqa
This coupling  reproduces exactly the second term in \reef{high}.  Since the above combination appears naturally in the string theory side, one may expect there there should be no on-shell ambiguity for the couplings in \reef{hderv2} if there is no such ambiguity for the couplings in \reef{hderv}. The couplings in \reef{hderv2} are the higher derivative extension of the coupling in the first term in the third line of \reef{exp2}.

To examine the other terms in the string theory amplitude \reef{pn}, consider the  expansion of $(t+1/4)J$ around \reef{point}, \ie
\beqa
(t+1/4)J&=&\frac{1}{2}\sqrt{2\pi}\left(\frac{-1}{(s+1/4)(t+s+u+1/2)}+ \frac{\sum_{n=0}^{\infty}a_n(s+t+u+1/2)^n}{(s+1/4)}\right.\labell{high2}\\
&&\left.+\frac{\sum_{n,m=0}^{\infty}d_{n,m}(s+t+1/2)^n(t+1/4)^{m+1}(s+1/4)^m}{(t+s+u+1/2)}\right.\nonumber\\
&&\left.+\sum_{p,n,m=0}^{\infty}e_{p,n,m}(s+t+u+1/2)^p(s+t+1/2)^n(t+1/4)^{m+1}(s+1/4)^m\right)\nonumber\eeqa
Replacing it in the amplitude \reef{pn},   one finds that  the first  term of \reef{high2} is reproduced by the effective actions  \reef{exp1} and \reef{exp2}.   The  second term of \reef{high2} on the other hand  should be reproduced by the following Feynman amplitude in field theory:
\beqa 
{\cal A}&=&V(C_{p-1},T_1,T_2)G(T_2)V(T_2,T_1,A^{(1)})\labell{finalA}
\eeqa
where the propagator and the vertex $V(T_2,T_1,A^{(1)})$ which have no higher derivative corrections are  given by 
\beqa
V(T_2,T_1,A^{(1)})&=&T_p(2\pi\alpha')(k_3-k)\inn\xi\nonumber\\
G(T_2)&=&\frac{i}{(2\pi\alpha')T_p(s+1/4)}\labell{ver2}\eeqa
and 
 the vertex $V(C_{p-1},T_1,T_2)$ should be derived from the higher derivative terms \reef{hderv}, that is
\beqa
V(C_{p-1},T_1,T_2)&=&(\alpha')^2\mu_p\sum_{n=0}^{\infty}a_n\left(-\frac{\alpha'}{2}p^ap_a\right)^n\frac{1}{p!} \eps^{a_{0}\cdots a_{p-1}a}H_{a_{0}\cdots a_{p-1}}k_{2a}\nonumber\eeqa
where $p^a$ is the momentum of the RR field. Replacing them in \reef{finalA}, one finds exact agreement with the second term in \reef{high2}. Note that the couplings \reef{hderv}  appears in \reef{finalA} as tachyonic pole and as contact term in  \reef{cont1}. Moreover, the combination of the tachyonic poles and the contact terms is \beqa
\eps^{a_{0}\cdots a_{p-1}a}H_{a_{0}\cdots a_{p-1}}\sum_{a=0}^{\infty}a_n(s+t+u+1/2)^n\left(\frac{2k_{2a}\xi\inn k_{3a}}{s+1/4}+\frac{2k_{3a}\xi\inn k_{2a}}{t+1/4}-\xi_a\right)\labell{gaugein}\eeqa
which satisfies the Ward identity. This  indicates that there is   no on-shell ambiguity in the couplings \reef{hderv}. We will discus it more in the Discussion section.

The sum of the massless poles in the second line of  \reef{high2} and the corresponding term when $k_2\leftrightarrow k_3$, and the massless pole in the second line of  \reef{high} is
\beqa
&&4i\mu_p\frac{ \eps^{a_{0}\cdots a_{p-1}a}H_{a_{0}\cdots a_{p-1}}}{p!(s+t+u+1/2)}
\left[k_{2a}(t+1/4)(2\xi.k_{3})
-
\frac{1}{2}\xi_a(s+1/4)(t+1/4)+(3\leftrightarrow 2)\right]\nonumber\\
&&\times \sum_{n,m=0}^{\infty}d_{n,m}(s+t+1/2)^n((t+1/4)(s+1/4))^m\labell{masslesspole}\eeqa
which satisfies the Ward identity. This should be reproduced in field theory by the amplitude \reef{amp4} in which the vertex $V_a(C_{p-1},A)$ and the propagator $G_{ab}(A)$ are given in \reef{amp4'} and the vertex $ V_b(A,A^{(1)},T_1,T_1)$ should be derived from the  the tachyon DBI action and its higher derivative extension in which we are not interested in this paper. In fact it has been checked in \cite{Garousi:2007fk} that the $d_{0,0}$ term is reproduced by the tachyon DBI action. 

Finally, the sum of the contact terms  in the last line of  \reef{high2} and the corresponding term when $k_2\leftrightarrow k_3$, and the last term in  \reef{high} is 
\beqa
&&4i\mu_p\frac{ \eps^{a_{0}\cdots a_{p-1}a}H_{a_{0}\cdots a_{p-1}}}{p!}
\left[k_{2a}(t+1/4)(2\xi.k_{3})
-
\frac{1}{2}\xi_a(s+1/4)(t+1/4)+(3\leftrightarrow 2)\right]\nonumber\\
&&\times \sum_{p,n,m=0}^{\infty}e_{p,n,m}(s+t+u+1/2)^p(s+t+1/2)^n((t+1/4)(s+1/4))^m\labell{contactterm}\eeqa
which satisfies the Ward identity. The field theory couplings corresponding to the above terms are
\beqa
2(\alpha')^2\mu_p\frac{ \eps^{a_{0}\cdots a_{p-1}a}H_{a_{0}\cdots a_{p-1}}}{p!}
\sum_{p,n,m=0}^{\infty}e_{p,n,m}(s+t+u+1/2)^p(s+t+1/2)^n((t+1/4)(s+1/4))^m\nonumber\\
\!\!\!\times \left[\prt_b\prt_c(A^{(1)}-A^{(2)})_aD^bT_1D^cT_1+2D_aD_bT_1D_cT_1\prt^b(A^{(1)}-A^{(2)})^c+T_1\rightarrow T_2\right]\nonumber\eeqa
where our notation is such that
\beqa
((s+1/4)(t+1/4))^mHATT&\rightarrow&\left(\alpha'\right)^{2m}H\prt_{a_1}\cdots\prt_{a_{2m}}AD^{a_1}\cdots D^{a_m}TD^{a_{m+1}}\cdots D^{a_{2m}}T\nonumber\\
(s+t+1/2)^nHATT&\rightarrow&\left(\alpha'\right)^nH\prt^{a_1}\cdots \prt^{a_n}AD_{a_1}\cdots D_{a_n}(TT)\nonumber\\
(s+t+u+1/2)^pHATT&\rightarrow&\left(\frac{\alpha'}{2}\right)^pH(D_aD^a)^p(ATT)\eeqa
Note that the above Lagrangian   is invariant under gauge transformation. In terms of field strength, these higher derivative couplings are
\beqa
&&2(\alpha')^2\mu_p
\sum_{p,n,m=0}^{\infty}e_{p,n,m}(s+t+u+1/2)^p(s+t+1/2)^n((t+1/4)(s+1/4))^m\nonumber\\
&& C_{p-1}\wedge \left[-\prt_b\prt_c(F^{(1)}-F^{(2)})D^bTD^cT^*+2D_bDT\wedge D_cDT^*(F^{(1)}-F^{(2)})^{bc}+\right.\nonumber\\
&&\left.+\prt_b(F^{(1)}-F^{(2)})_{c}\wedge D^bDTD^cT^*+\prt_b(F^{(1)}-F^{(2)})_{c}\wedge D^bDT^*D^cT \right]\labell{highpn'}\eeqa
Our notation is that the fields without  indexes are forms, \eg $F^{(1)}_c$ is one form and $F^{(1)}$ is two form. The above couplings have on-shell ambiguity which can be fixed by studying the S-matrix element of one RR, two tachyons and two gauge fields in which the above couplings appear in the tachyonic pole and in  the contact terms of the amplitude. The above higher derivative couplings have at least four derivatives, so they are not extension of the couplings in \reef{exp2}. This ends our illustration of consistency between the expansion of the S-matrix element of one RR, two tachyons and one gauge field around \reef{point} and the higher derivative couplings of the field theory.

\section {Discussion}

In this paper, we have shown that the  expansion of the S-matrix element of one RR, two tachyons and one gauge field around   \reef{point} corresponds to higher derivative extension of the Wess-Zumino terms, \ie the couplings \reef{hderv12}, \reef{hderv2} and \reef{highpn'}. Hence, one expects that the expansion  \reef{point} to be the momentum expansion, \ie an expansion that its leading order terms correspond to the effective action and its non-leading terms correspond to the higher derivative extension of the effective action.  In fact, the expansion \reef{point} in terms of momenta of the open string fields is 
\beqa
\alpha'k_1\cdot k_2\rightarrow 0, \, \alpha'k_1\cdot k_3\rightarrow 0, \, \alpha'(k_2+ k_3)^2\rightarrow 0
\eeqa
The low energy expansion of the amplitude \reef{amp33} is also around
\beqa
\alpha'( k+ k')^2\rightarrow 0\eeqa
To find a general rule for the momentum expansion of any S-matrix element involving tachyon, let us examine the expansion of some other S-matrix elements.  The  momentum expansion of the S-matrix element of two gauge fields and two tachyons  has been proposed in \cite{Garousi:2002wq} to be  around
\beqa
\alpha'k_1\cdot k_2\rightarrow 0, \, \alpha'k_1\cdot k_3\rightarrow 0, \, \alpha'k_2\cdot k_3\rightarrow 0\labell{ttaa}
\eeqa
where $k_1$ is momentum of tachyon and $k_2,\, k_3$ are the momenta of the gauge fields. The momentum expansion of the S-matrix element of four tachyons has been proposed  in \cite{Garousi:2002wq}. The amplitude has different Chan-Paton factors. The one which has the factor $\Tr(\l_1\l_2\l_3\l_4)$ should be expanded around 
\beqa
&&\left(\alpha'(k_1+k_2)^2,\, \alpha'k_2\cdot k_3,\,\alpha'k_1\cdot k_3\right)\rightarrow 0\nonumber\\
&&-\left(\alpha'(k_1+k_3)^2,\, \alpha'k_2\cdot k_3,\,\alpha'k_1\cdot k_2\right)\rightarrow 0\nonumber\\
&&+\left(\alpha'(k_2+k_3)^2,\, \alpha'k_1\cdot k_2,\,\alpha'k_1\cdot k_3\right)\rightarrow 0\nonumber
\eeqa
The first line produce $s$-channel, the second line produces $u$-channel and the last one produces the $t$-channel. The momentum  expansion of the S-matrix element of four tachyons and one gauge field has been proposed in \cite{BitaghsirFadafan:2006cj}. The amplitude has different Chan-Paton factors and different factors of $k_1\cdot\z_5$, $k_2\cdot\z_5$ and $k_3\cdot\z_5$ where $\z_5$ is the polarization of the gauge field. The one which has the factor $\Tr(\l_1\l_2\l_3\l_4\l_5)k_1\cdot\z_5$ should  be expanded around 
\beqa
&&\left(\alpha'(k_1+k_2)^2,\,\alpha'(k_3+k_4)^2, \, \alpha'k_2\cdot k_3,\,\alpha'k_1\cdot k_5,\,\alpha'k_4\cdot k_5\right)\rightarrow 0\nonumber\\
&&-\left(\alpha'k_1\cdot k_2,\,\alpha'k_3\cdot k_4, \, \alpha'k_2\cdot k_3,\,\alpha'k_1\cdot k_5,\,\alpha'k_4\cdot k_5\right)\rightarrow 0\nonumber\\
&&+\left(\alpha'(k_2+k_3)^2,\,\alpha'k_3\cdot k_4, \, \alpha'k_2\cdot k_3,\,\alpha'k_1\cdot k_5,\,\alpha'k_4\cdot k_5\right)\rightarrow 0\nonumber
\eeqa
Let us compare the above expansions  with  the momentum expansion of the S-matrix elements involving only massless fields. The momentum expansion in this case  is trivial, \ie expansion around  $\alpha' k_i\cdot k_j\rightarrow 0$ which is equivalent to the expansion around $\alpha'(k_i+k_j)^2\rightarrow 0$. However, for tachyon the expansion around $\alpha' k_i\cdot k_j\rightarrow 0$ is not equivalent to the expansion around $\alpha'(k_i+k_j)^2\rightarrow 0$. For example, the expansion of amplitude \reef{amp33} around $\alpha'k\cdot k'\rightarrow 0$ does not produce the massless pole of field theory, hence, that expansion would not be  correspond to the effective field theory of brane-anti-brane system. In general, the  momentum expansion of a tachyon amplitude should be an  expansion around $\alpha' k_i\cdot k_j\rightarrow 0$ or $\alpha'(k_i+k_j)^2\rightarrow 0$ or a combination of them for each $i,j$.
The nontrivial question of finding the  momentum expansion of a tachyon amplitude  in then correspond to fixing this ambiguity.

The above expansions of the tachyon amplitudes are then the expansions in terms of power of momenta of the external states, \ie $\alpha'$ expansions. Hence, one expects that they  should be correspond to the derivative expansions of the field theory. In particular,  expansion of the S-matrix element of two tachyons and two gauge fields around \reef{ttaa} should be correspond to the higher derivative  coupling of two tachyons and two gauge fields. These higher derivative couplings on the other hand can be used to find the vertex $V_b(A,A^{(1)},T_1,T_1)$ in the amplitude \reef{amp4} to produce the massless pole in \reef{masslesspole}. The details of these calculations will be reported in a forthcoming  paper\cite{Garousi:2008xp}, we note here  that the constants $d_{n,m}$ that appear in the amplitude \reef{masslesspole} have the same structure as the coefficients of expansion of the S-matrix element of two gauge fields and two tachyons \cite{Garousi:2002wq} expanded around \reef{ttaa}, in particular they do not have $\ln(2)'s$. The comparison of the field theory massless pole with the string theory may also give some residual contact terms. These contact terms would have the same structure as those appear in \reef{contactterm}, so they would modify the coefficients $e_{p,n,m}$ in \reef{highpn'}. 

The above rule for expanding the open string tachyon amplitude should be hold even for closed string tachyon amplitude. The expansion of the sphere level S-matrix element of two gravitons and two closed string tachyons in type 0 theory has been proposed  in \cite{Garousi:2003db} to be around
\beqa
\alpha'(p_3+p_4)\rightarrow 0,\,\alpha'p_1\cdot p_4\rightarrow 0,\, \alpha'p_2\cdot p_4\rightarrow 0\nonumber\eeqa
where $p_1,\, p_2$ are the graviton momenta and $p_3,\, p_4$ are the tachyon momenta.  The expansion of the  S-matrix element of four closed string tachyons in type 0 theory has been proposed in \cite{Garousi:2003db} to be around
\beqa
&&\frac{1}{3}\left( [\alpha'(p_3+p_4),\,\alpha'p_1\cdot p_4,\, \alpha' p_2\cdot p_4]\rightarrow 0\right.\nonumber\\
&&+\left.[\alpha'(p_1+p_4),\,\alpha'p_3\cdot p_4,\, \alpha' p_2\cdot p_4]\rightarrow 0\right.\nonumber\\
&&+\left.[\alpha'(p_2+p_4),\,\alpha'p_3\cdot p_4,\, \alpha' p_1\cdot p_4]\rightarrow 0\right)
\eeqa
which correspond to $s$-, $t$- and $u$-channels. 

The strategy for finding the above expansions in \cite{Garousi:2003db} was to find an expansion whose leading order terms are reproduced exactly by the  effective  action of type 0 theory which includes a covariant tachyon kinetic term.  Using this strategy  along for finding the expansion of the S-matrix element of two RR and two tachyons, one does not find a unique expansion. In fact in \cite{Garousi:2003db} three expansions has been examined  which are consistent with the effective field theory.  However, neither of them is consistent with the above rule for finding the  momentum  expansion. A momentum  expansion which is also consistent with the strategy in \cite{Garousi:2003db}  is $(s\rightarrow 0,\, t,u\rightarrow -1)/2+( s\rightarrow -2,\, t,u\rightarrow 0)/2$ which in terms of momentum is
\beqa
&&\frac{1}{2}\left( [\alpha'(p_3+p_4),\,\alpha'p_1\cdot p_4,\, \alpha' p_2\cdot p_4]\rightarrow 0\right.\nonumber\\
&&+\left.[\alpha'(p_1+p_4),\,\alpha'(p_2+p_4), \,\alpha'p_3\cdot p_4,]\rightarrow 0\right)
\eeqa
The expansion of the amplitude around this point is \beqa
A(C,C,T,T)&\sim& \pi\alpha\left(-\frac{2}{s}-\frac{2}{t}-\frac{2}{u}+4\ln(2)+\cdots\right)\eeqa
where $\alpha$ is a factor which has four momenta. The above expansion is consistent with the effective action of type 0 theory and fixes the function $f(T)$ that multiply the kinetic term of the RR fields to be 
\beqa
f(T)&=&1+T+\frac{1}{2}T^2 \nonumber\eeqa
which is  the one that has been found in \cite{Klebanov:1998yya}.

We have seen that infinite higher derivative couplings in \reef{hderv} which have been read from the momentum expansion of the S-matrix element of one RR and two tachyons, appear as tachyonic pole of the momentum expansion of the S-matrix element of one RR, two tachyons and one gauge field. This indicates that there should be only one  momentum expansion for the string theory   S-matrix element of one RR, two tachyons and one gauge field which is consistent with the field theory, \ie if one expands the amplitude \reef{gen} around a point other than \reef{point}, one would not find the higher derivative couplings \reef{hderv} in the tachyonic   pole of the amplitude \reef{gen}. This consistency  between different S-matrix element should be holed for all other S-matrix elements. Therefore, one expects that for any string theory S-matrix element there should be a unique momentum expansion, \eg the above expansion points should be unique.

The tachyon couplings \reef{hderv} appear as tachyonic pole and the contact terms of string theory S-matrix element \reef{gen} with exactly correct coefficients. The sum of these two terms is gauge invariant. We have interpreted this as an indication  that the couplings in \reef{hderv} have no on-shell ambiguity. To elaborate  this point, consider, as an example, the following couplings: 
 \beqa
2i\alpha'\mu_p\sum_{n=0}^{\infty}a_n\left(\frac{\alpha'}{2}\right)^n(-2\alpha')^2 C_{p-1}\wedge (D ^aD_a)^n(DD_{\alpha}D^{\alpha}T\wedge DD_{\beta}D^{\beta}T^*)\labell{hderv0}
\eeqa
which is equivalent to \reef{hderv} for on-shell tachyon. If one considers the above couplings instead of the couplings \reef{hderv}, one would find the contact terms in the second term in \reef{high} after combining them  with the couplings in \reef{hderv2}. The above couplings produce also  the tachyonic pole \reef{finalA} in which the vertex $V(C_{p-1},T_1,T_2)$ is
 \beqa
V(C_{p-1},T_1,T_2)&=&(\alpha')^2\mu_p\sum_{n=0}^{\infty}a_n\left(-\frac{\alpha'}{2}p^ap_a\right)^n\frac{1}{p!} \eps^{a_{0}\cdots a_{p-1}a}H_{a_{0}\cdots a_{p-1}}k_{2a}(2\alpha'k^2)\nonumber\eeqa 
where $k=k_1+k_3$ is the momentum of the off-shell tachyon. If one replaces it into \reef{finalA}, one finds  an extra factor of $(2\alpha'k^2)$ in the tachyonic pole. However, one can write it as
\beqa
2\alpha'k^2&=&-4s\,=\,1-4(s+\frac{1}{4})\nonumber\eeqa
the first term gives exactly the tachyonic pole which is also produced by \reef{hderv}. When combining it with the above contact terms one finds the gauge invariant combination \reef{gaugein}.  The second term on the other hand gives an extra contact term. The resulting contact terms are
\beqa
\eps^{a_{0}\cdots a_{p-1}a}H_{a_{0}\cdots a_{p-1}}\sum_{n=0}^{\infty}a_n(s+t+u+1/2)^n(\xi\inn k_3\,k_{2a}+\xi\inn k_2\,k_{3a})\eeqa
Obviously it does not satisfy the Ward identity, so it can not be reproduced by a gauge invariant coupling in field theory. Therefore, one would find inconsistency between field theory and string theory S-matrix element if one considers the couplings \reef{hderv0} instead of the couplings \reef{hderv}. We expect similar idea should be hole for all other tachyon couplings. In other words, the S-matrix method in principle may have  the potential  to produce all tachyon couplings without on-shell ambiguity.  
%%%%%%%%%%%%%%%%%%%%%%%%%%%%%%%%%%%%%%%%%%%%%%%%%%%%%%%%%%%%%%%%%%%%%%%%%%%%%%%%%%%%%%%
%%%%%%%%%%%%%%%%%%%%%%%%%%%%%%%%%%%%%%%%%%%%%%%%%%%%%%%%%%%%%
%Appendixes\renewcommand{\thesection}{A}
  %%%%%%%%%%%%%%%%%%%%%%%%%%%%%%%%%%%%%%%%%%%%%%%%%%%%%%%%%%%%%%%%%%%%%%%%%%%%%%%%%%%%%%%
\section*{Acknowledgment}

I would like to thank A. Ghodsi for  discussion .

%%%%%%%%%%%%%%%%%%%%%%%%%%%%%%%%%%%%%%%%%%%%%%%%%%%%%%%%%%%%%%%%%%%%%%%%%%%%%%%%%%%%%%%

\end{document}